\documentclass[a4paper]{article}

\usepackage{INTERSPEECH2020}

\usepackage{enumitem}
\usepackage{multirow,comment,xspace}
\usepackage{array}

\newcommand{\rcp}[1]{\textcolor{red}{Ref Ch. P.}}

\usepackage{setspace,url}
\usepackage{lineno,multirow,cite}
\usepackage{authblk}
\usepackage{epsfig,amssymb}
\usepackage[utf8]{inputenc}
\usepackage{amsmath,graphicx}
\usepackage{amssymb,epsfig,url,mathrsfs} 
\usepackage{xcolor}
\usepackage{multirow}
\usepackage{multicol}
\usepackage{enumitem}
\usepackage{dblfloatfix}

\usepackage{numprint}
\npthousandsep{,}

\usepackage{graphicx}
\usepackage{array, makecell}
\usepackage{verbatim}

\usepackage[tight]{subfigure}
\usepackage[utf8]{inputenc}
\usepackage[T1]{fontenc}
\usepackage{tabularx}  
\usepackage{ragged2e}  
\newcolumntype{Y}{>{\RaggedRight\arraybackslash}X} 
\usepackage{booktabs}  
\PassOptionsToPackage{hyphens}{url}
\usepackage[bookmarks=false,hidelinks]{hyperref}

\title{Design Choices for X-vector Based Speaker Anonymization}
\name{Brij Mohan Lal Srivastava$^1$, Natalia Tomashenko$^2$, Xin Wang$^3$, Emmanuel Vincent$^4$,\\ Junichi Yamagishi$^3$, Mohamed Maouche$^1$, Aur\'elien Bellet$^1$, Marc Tommasi$^5$}
\address{
  $^1$Inria, France $^2$Laboratoire Informatique d’Avignon (LIA), Avignon Université, France\\
  $^3$National Institute of Informatics, Tokyo, Japan\\
  $^4$Université de Lorraine, CNRS, Inria, LORIA, France
  $^5$Universit\'e de Lille, France}
\email{organisers@lists.voiceprivacychallenge.org}

\begin{document}

\maketitle
\begin{abstract}
 
 The recently proposed x-vector based anonymization scheme converts any input voice into that of a random {\it pseudo-speaker}. In this paper, we present a flexible
 pseudo-speaker selection technique as a baseline
 for the first VoicePrivacy Challenge. We explore several design choices for the distance metric between speakers, the
 region of x-vector space where the pseudo-speaker is picked, and gender
 selection. To assess the strength of anonymization achieved,
 we consider attackers using an x-vector based speaker verification
 system who may use original or anonymized speech for enrollment, depending on
 their knowledge of the anonymization scheme. 
 The Equal Error Rate (EER) achieved by the attackers and the decoding Word Error Rate (WER) over anonymized data are reported as the measures
 of privacy and utility. Experiments are
 performed using datasets derived from LibriSpeech to find the optimal combination of design choices in terms of privacy and utility.
 
\end{abstract}


\section{Introduction}

Privacy protection methods for speech fall into four broad
categories \cite{tomashenko:hal-02562199}: deletion, encryption, distributed learning, and anonymization. The VoicePrivacy
initiative \cite{tomashenko:hal-02562199} specifically promotes the development of \textit{anonymization} methods which aim to
suppress personally identifiable information in speech while leaving
other attributes such as linguistic content intact.\footnote{In the legal community, the term ``anonymization'' means that this goal has been achieved. Following the VoicePrivacy Challenge, we use it to refer to the task to be addressed, even when the method has failed.} Recent studies have proposed anonymization methods
based on noise addition \cite{hashimoto2016privacy}, speech
transformation \cite{qian2017voicemask}, voice conversion~
\cite{jin2009speaker,pobar2014online,bahmaninezhad2018convolutional}, speech
synthesis \cite{justin2015speaker,fang2019speaker}, and adversarial
learning \cite{srivastava2019privacy}. 
We focus on voice conversion / speech synthesis based methods due to the naturalness
of their output and their promising results so far.

In order to implement a speaker anonymization scheme based on voice
conversion or speech synthesis, we must address the following questions: 1. \textit{What is the best
representation to characterize speaker information in a speech signal?} 2. 
\textit{Which distance metric is most appropriate to
explore various regions of the speaker space?} 3. \textit{How to optimally
select target speakers from a 
small pool of speakers?} 4. \textit{How to combine the distance metric and target selection in order to strike balance between privacy protection and loss of utility?}

Classically, speaker anonymization methods that rely on a voice conversion or
speech synthesis system select a random target speaker from a pool of speakers which must be included in the training set for that system. This constraint severely restricts the user's freedom to choose an arbitrary unseen speaker as the target for anonymization. Moreover, several targets cannot be mixed together to create an imaginary sample in speaker space, i.e., a {\it pseudo-speaker}. In a previous experimental study  \cite{srivastava2019evaluating}, we specified three criteria to be satisfied by voice conversion algorithms for speaker anonymization: 1) non-parallel, 2) many-to-many, and 3) source- and language-independent. Although the algorithms compared in \cite{srivastava2019evaluating} satisfied these criteria, they did not allow conversion conditioned over a continuous speaker representation, such as x-vectors \cite{snyder2018x}.


Recently, Fang et al.\ \cite{fang2019speaker} proposed to identify x-vectors at a
fixed distance from the ``user'' x-vector and to combine them to produce a 
{\it
pseudo-speaker} representation. This representation, along with the ``user''
linguistic representation, is provided as input to a Neural Source-Filter 
(NSF) \cite{wang2019neural1} based speech synthesizer to produce
anonymized speech. Han et al.\ \cite{han2020voice}
extended \cite{fang2019speaker} by proposing a metric privacy framework where an
x-vector based \textit{pseudo-speaker} is selected so as to satisfy a given
privacy budget. Based on these studies, we answer Question
1 by choosing x-vectors as the appropriate
speaker representation.
In addition, the freedom to generate previously unseen pseudo-speakers by combining existing speakers from a 
small dataset exponentially increases the choices for the user.

The user may select pseudo-speakers at random in the entire
x-vector space or based on specific properties, such as density of speakers, gender majority, etc. They must also choose a similarity metric between x-vectors since this dictates the properties of the vector space.
Previous studies \cite{kenny2010bayesian} have shown
that Probabilistic Linear Discriminant Analysis (PLDA) yields state-of-the-art
speaker verification performance, superior to the cosine distance.
This is attributed to the formulation of PLDA
which estimates the factorized {\it within-speaker} and {\it between-speaker} variability in speaker space. Hence, the PLDA score provides a good estimate of the log-likelihood ratio between {\it same-speaker} and {\it different-speaker} hypotheses, making it a superior measure of speaker affinity even for short speech segments \cite{salmun2016use}. 

In this paper, we establish that a greater level of anonymization is
achieved when the distance between x-vectors is measured by PLDA
instead of the cosine distance as used by
Fang et al.\ \cite{fang2019speaker} (answering Question 2). Then, we introduce a
design choice called {\it proximity} which allows us to pick the pseudo-speaker
in {\it dense}, {\it sparse}, {\it far}, or {\it near} regions of speaker space.
We further explore the flexibility of this anonymization scheme by
exploring the influence of gender selection. 
These design choices are evaluated using attackers which may or
may not know the anonymization scheme applied (answering
Question 3).
Finally we suggest the optimal combination of distance metric and design choices
based on qualitative and quantitative measures to balance privacy and utility (answering Question 4).

We describe the general anonymization framework and the proposed design choices in Section~\ref{sec:anon-framework}. The datasets and evaluations metrics are briefly explained in Section~\ref{sec:exp-setup}. We present the experiments and discuss their results in Section~\ref{sec:exp-results}. Section~\ref{sec:conc} concludes the paper.

\section{Anonymization design choices}
\label{sec:anon-framework}

\begin{figure}[h!]
\centering
\includegraphics[width=\linewidth]{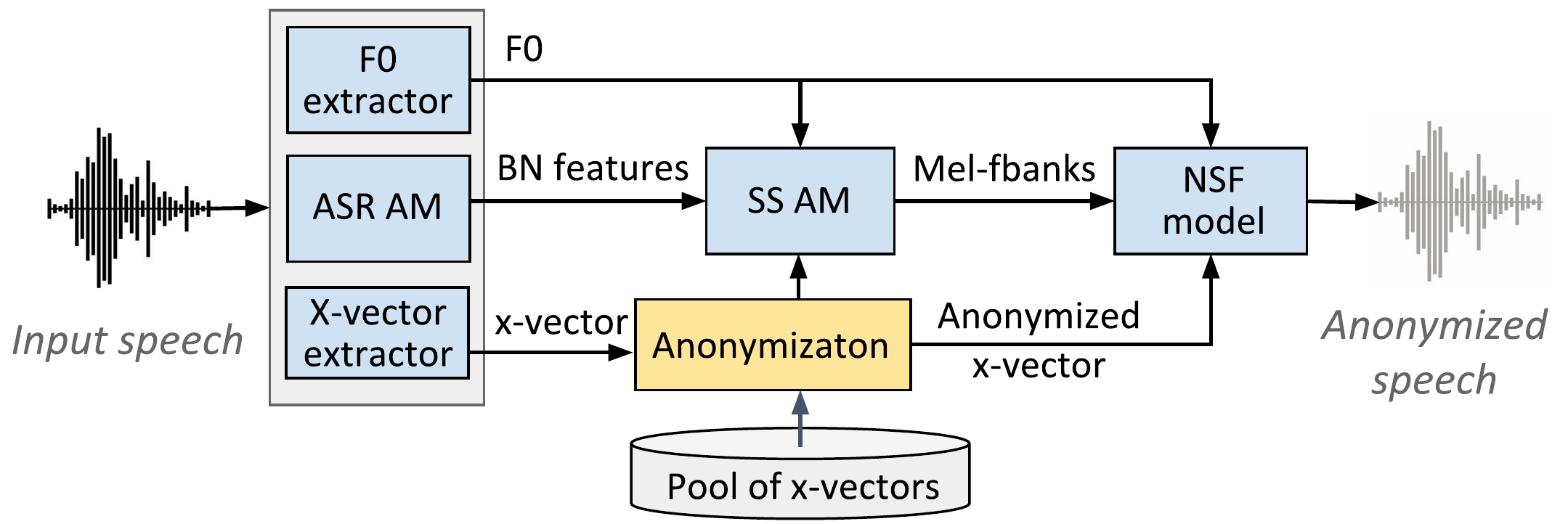}
\caption{General anonymization scheme.}
\label{fig:baseline}
\end{figure}

The general anonymization scheme follows the method proposed
in \cite{tomashenko2020voiceprivacy} and shown in Fig.~\ref{fig:baseline}.
It comprises three steps: \textit{Step 1 (Feature
extraction)} extracts fundamental frequency (F0) and bottleneck (BN)
features and the source speaker's x-vector from the input signal. \textit{Step 2 (X-vector anonymization)} anonymizes this x-vector using an external pool of speakers. \textit{Step 3 (Speech synthesis)} synthesizes a speech waveform from the anonymized x-vector and the original BN and F0 features using an acoustic model (AM) and the NSF model.

%

\textsl{Step 2} (yellow box in Fig.~\ref{fig:baseline})
is the focus of this paper. It aims to generate a pseudo-speaker and comprises two sub-steps: 1) select $N^*$ candidate target x-vectors from the {\it anonymization pool}; 2) average them
to obtain the {\it pseudo-speaker} x-vector. 
In the following, we introduce various design choices for pseudo-speaker
selection.
In all cases, a single target pseudo-speaker x-vector is selected for a given
source speaker $S$, and all the utterances of $S$ are mapped to it, following
the {\it perm} strategy described in \cite{srivastava2019evaluating}.
This strategy has been shown to perform robust anonymization compared to other
strategies described in \cite{srivastava2019evaluating}.

\subsection{Distance Metric: Cosine vs.\ PLDA}
\label{sec:dc_distance}

We compare two metrics to identify candidates for target x-vectors.
The first one is the cosine distance, which was used
by \cite{fang2019speaker}. It is defined as
\begin{equation}
1-\frac{u \cdot v}{||u||_2
||v||_2}
\end{equation}
for a pair of x-vectors $u$ and $v$. The second one is
PLDA \cite{ioffe2006probabilistic}, which represents the log-likelihood ratio
of {\it same-speaker} ($\mathcal{H}_s$) and {\it different-speaker} ($\mathcal{H}_d$) hypotheses.
PLDA models x-vectors $\omega$ as $\omega = m + Vy + Dz$, where $m$ is the center of the acoustic space, the columns of $V$ represent speaker variability (eigenvoices) with $y$ depending only on the speaker, and the columns of $D$ capture channel variability (eigenchannels) with $z$ varying from one recording to another.
The parameters $m$, $V$ and $D$ are trained using x-vectors from the training set for the x-vector model,
which is used to generate the {\it anonymization pool}. 
The log-likelihood ratio score
\begin{equation}
\text{PLDA}=\log \frac{p(\omega_i, \omega_j | \mathcal{H}_s)}{p(\omega_i, \omega_j | \mathcal{H}_d)}
\end{equation}
can be computed in closed form \cite{rohdin2014constrained}. 
We propose to use minus-$\text{PLDA}$ as the ``distance'' between a pair of x-vectors.

\subsection{Proximity: Random}
\label{sec:dc_proximity_random}

The simplest candidate x-vector selection strategy called \emph{random} consists of simply selecting $N^*$ (set to $100$) x-vectors uniformly at
random from
the same gender as the source in the {\it anonymization pool}. Note that this strategy does not allow us to choose particular regions of interest in x-vector space.

\subsection{Proximity: Near vs.\ Far}
\label{sec:dc_proximity_nf}

The notion of distance can be used to define regions in x-vector space which closely resemble ({\it near}) or least resemble ({\it far}) the source speaker $S$. In essence, we rank all the x-vectors in the anonymization pool in increasing order of their distance from $S$ and select either the top $N$ ({\it near}) or the bottom $N$ ({\it far}). To introduce some randomness, $N^* <N$ x-vectors are selected out of these $N$ uniformly at random. The variability of results is controlled by a fixed random seed. The values of $N$ and $N^*$ are fixed to 200 and 100 respectively in our experiments.
We noticed a sharp decline in utility for a smaller value of $N^*$.


\subsection{Proximity: Sparse vs.\ Dense}
\label{sec:dc_proximity_sd}

A simple mapping to {\it far} or {\it near} regions might produce biased {\it pseudo-speaker} estimates and the actual region where the output x-vector lies may not be optimal with respect to the distance from the source speaker. In order to pick the target {\it pseudo-speaker} in a specific region, we identify clusters of x-vectors in the {\it anonymization pool} which are then ranked based on their density. The density of each cluster is determined by the number of members belonging to that cluster.

We use Affinity Propagation \cite{dueck2009affinity} to
determine the number of clusters and their members in the {\it anonymization pool}.
Affinity Propagation is a non-parametric clustering method where the number of
clusters is determined automatically through a message passing protocol.
Two parameters determine the final number of clusters: {\it preference} assigns prior weights to samples which may be likely candidates for centroids, and {\it damping factor} is a floating-point multiplier to responsibility and availability messages. In our experiments, equal {\it preference} is assigned to each sample and the {\it damping factor} is set to 0.5. Out of 1160 speakers in the {\it anonymization pool}, 80 clusters were
found, including 46 male and 34 female. The number of speakers per cluster ranges from 6 
(sparse) to 36 (dense).

Candidate x-vector selection is achieved by picking either the 10 clusters with least members ({\it sparse}) or the 10 clusters with most members ({\it dense}). The remaining clusters are ignored. During anonymization, one of the 10 clusters is selected at random and 50\% of its members ($N^*$) are averaged to produce the {\it pseudo-speaker}. The 50\% candidate x-vectors for a given cluster remain fixed for a given random seed.

\subsection{Gender-selection: Same, Opposite, or Random}
\label{sec:dc_gs}

We observe clear clustering of the two genders in x-vector space using both
cosine and PLDA distances. Hence, we propose gender selection as a
design
choice to study its impact on anonymization and intelligibility.
We have the gender information for the source speaker as well as the speakers
in the {\it anonymization pool}. Hence this design choice can be combined with
all {\it proximity} choices. We study three different types of gender
selection: {\it same} where the candidate target x-vectors are constrained to be of the same gender as the source; {\it opposite} where they are constrained to be of the opposite gender; and {\it random} where the target gender is selected at random before picking candidate x-vectors of that gender.

\begin{figure*}[b]
\centering
  \subfigure[Distance]{\label{fig:eer_distance}\includegraphics[width=0.33\linewidth,trim=14 14 14 14,clip]{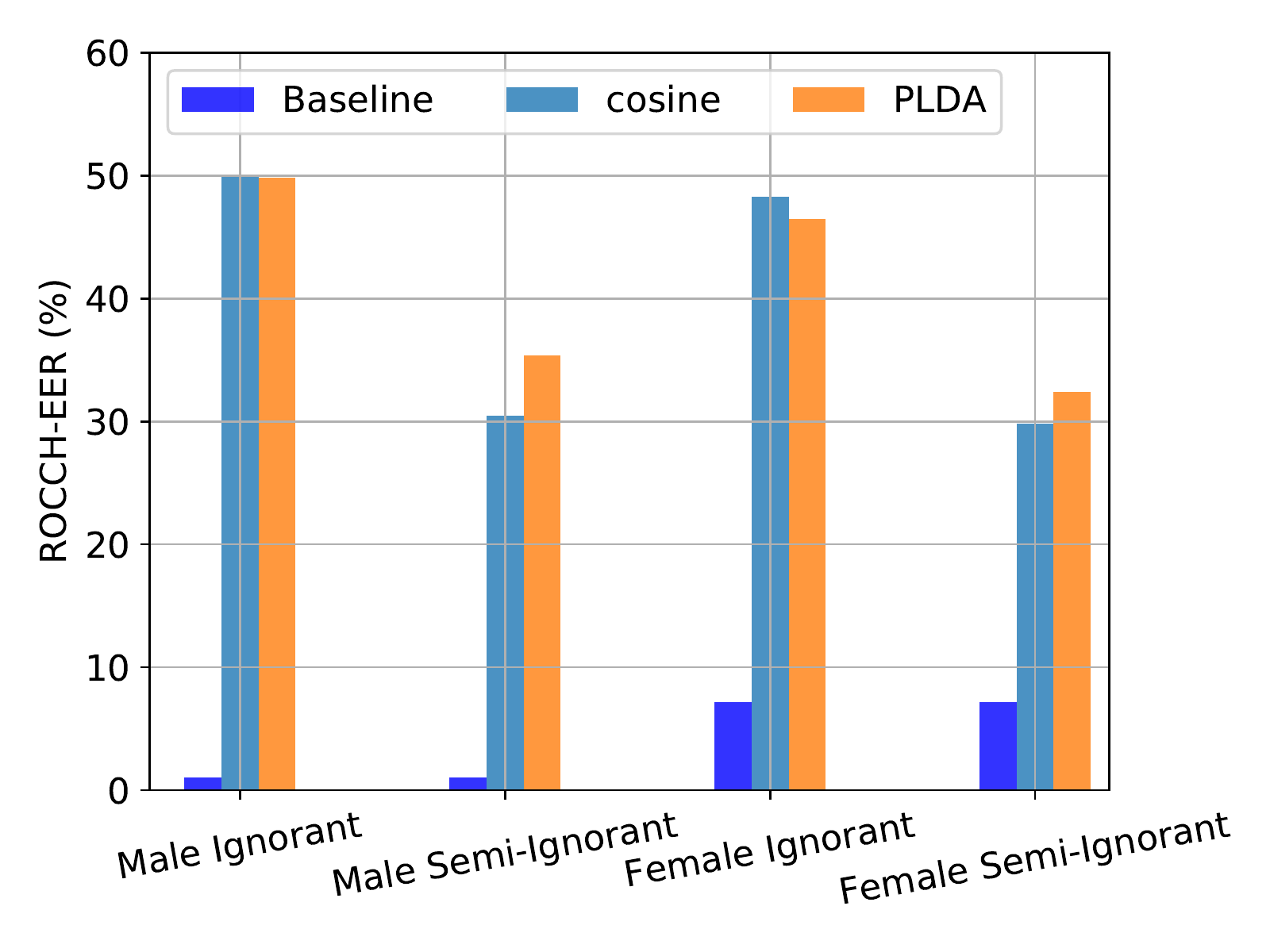}}
  \subfigure[Proximity]{\label{fig:eer_proximity}\includegraphics[width=0.33\linewidth,trim=14 14 14 14,clip]{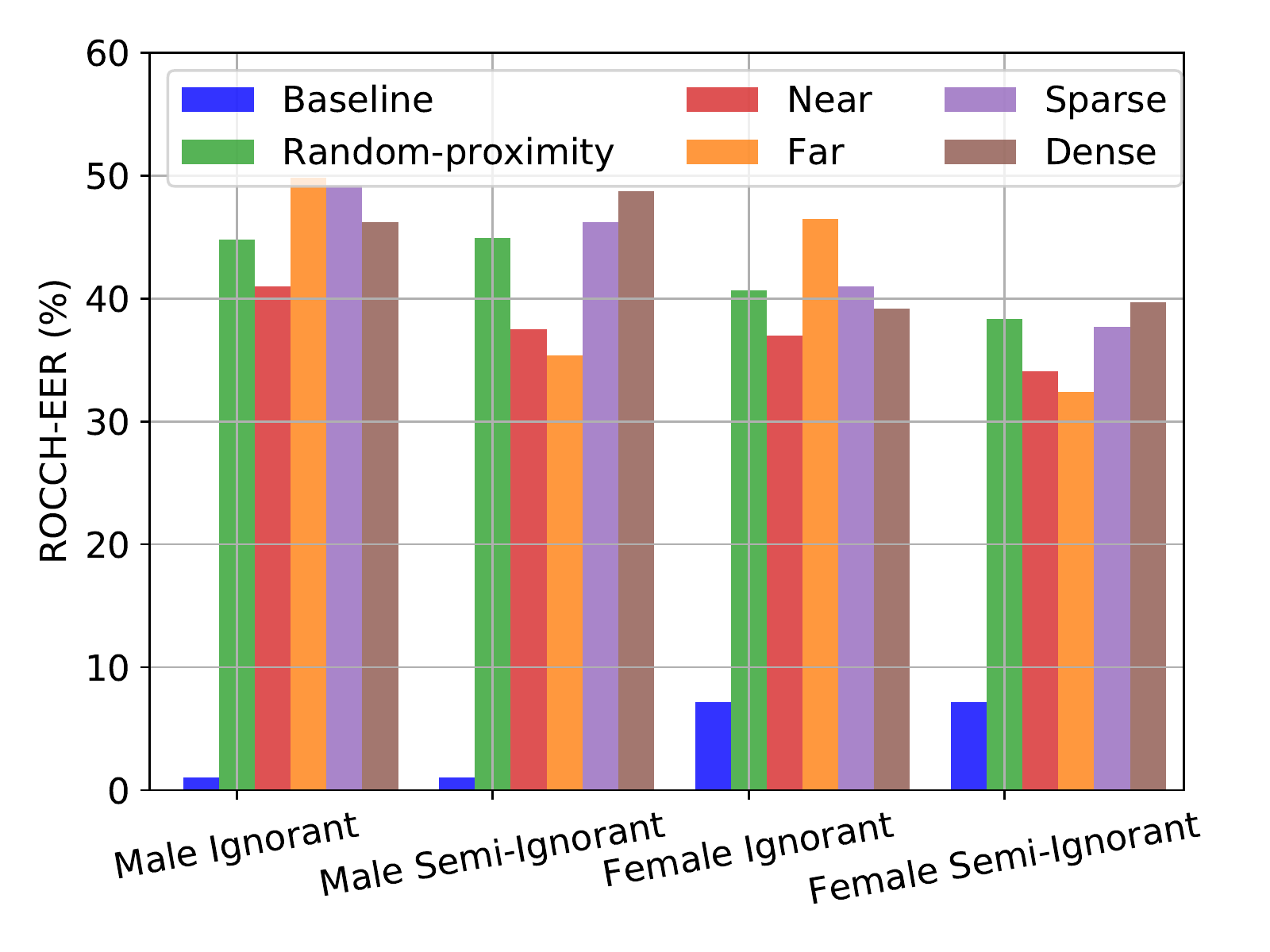}}
  \subfigure[Gender selection]{\label{fig:eer_gender}\includegraphics[width=0.33\linewidth,trim=14 14 14 14,clip]{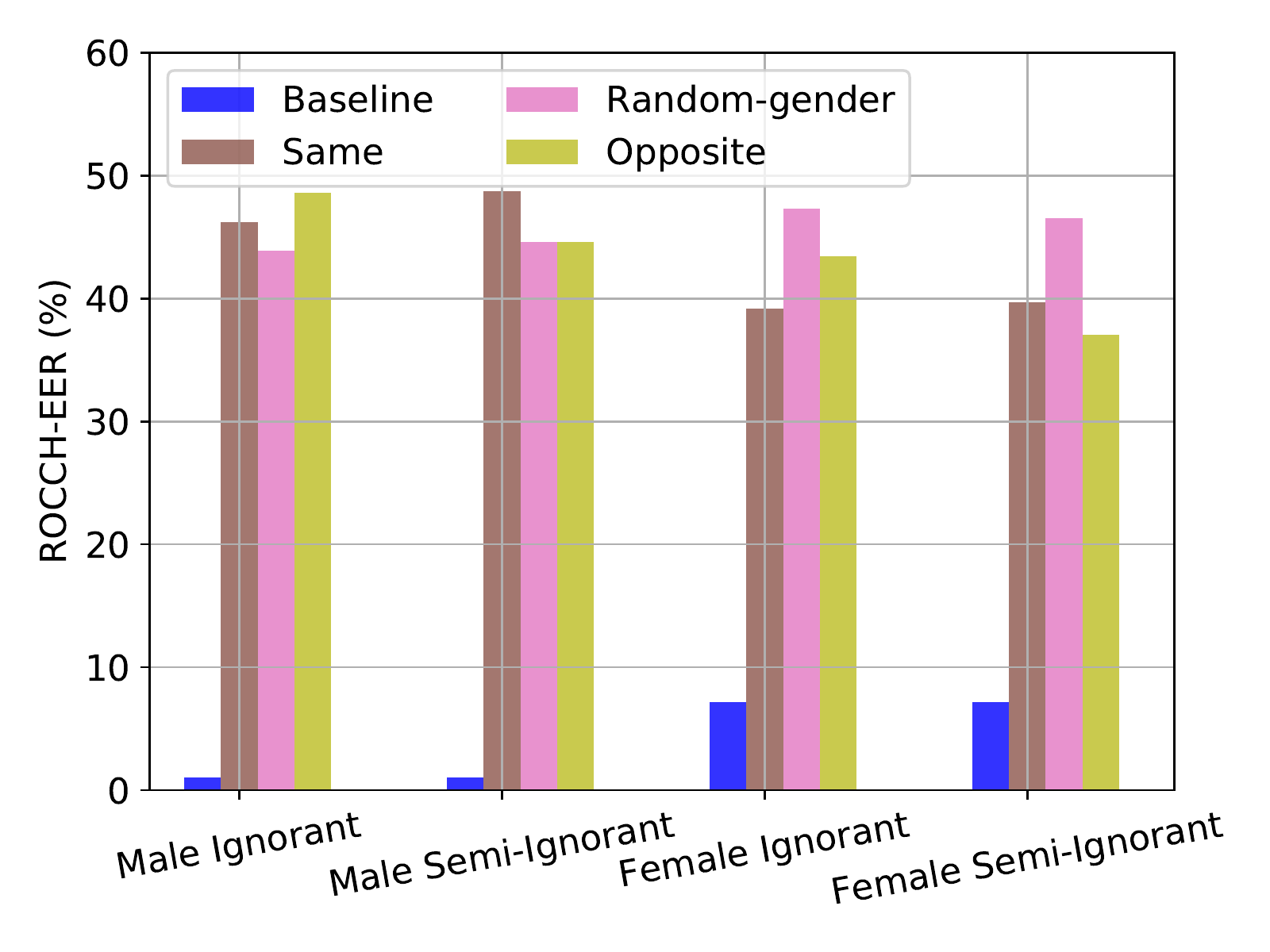}}
  
  \caption{ROCCH-EER (\%) obtained by \emph{$ASV_\text{eval}$} on the test set by an \emph{Ignorant} or a \emph{Semi-Ignorant} attacker for different design choices.  a)  Distance: \emph{cosine} vs.\ \emph{PLDA}. Proximity is fixed to \emph{far} and gender to \emph{same}. b) Proximity: \emph{random}, \emph{near}, \emph{far}, \emph{sparse}, or \emph{dense}. Distance is fixed to \emph{PLDA} and gender to \emph{same}. c) Gender: \emph{same}, \emph{opposite}, or \emph{random}. Distance is fixed to \emph{PLDA} and proximity to \emph{dense}.}
  \label{fig:eer_bars}
\end{figure*}

\section{Experimental setup}
\label{sec:exp-setup}

\subsection{Data}

Following the rules of the VoicePrivacy Challenge, we use three publicly available datasets for our experiments.\footnote{The VoicePrivacy Challenge involves development and evaluation sets built from both LibriSpeech and VCTK. Due to space limitations, we focus on LibriSpeech here.}
VoxCeleb-1,2 \cite{nagrani2017voxceleb,chung2018voxceleb2} and the \emph{train-clean-100} and \emph{train-other-500} subsets of LibriSpeech \cite{panayotov2015librispeech} and LibriTTS \cite{zen2019libritts} are used to train the models
described in Section~\ref{sec:anon-framework}. The development and test
sets are built from LibriSpeech {\it dev-clean} and 
{\it test-clean}, respectively. Details about the number of
speakers, utterances, and trials in the enrollment and trial
sets can be found in \cite{tomashenko:hal-02562199}.

\subsection{Evaluation methodology}

We evaluate the above design choices in terms of privacy and utility. We
define utility as the objective intelligibility of anonymized speech measured by the Word Error Rate (WER). The primary metric for privacy is the Equal Error Rate (EER).

\subsubsection{Attack model}

Privacy protection can be seen as a game between two entities: a ``user'' who
publishes anonymized speech to hide his/ her identity, and an ``attacker'' who
attempts to uncover the user's identity by conducting speaker verification
trials over enrolled speakers. The attacker may possibly use some knowledge about
the anonymization scheme to transform the enrollment data.

To assess the strength of anonymization against attackers with increasing amounts of knowledge, we perform the evaluation in three stages. The first
scenario ({\it Baseline}) refers to the case when the user does not perform
any anonymization before publication and the attacker also uses non-anonymized
speech for enrollment. This attacker typically achieves low error rate (i.e., the
user identity is accurately predicted) since there is no anonymization. In the
second scenario ({\it Ignorant}), the user publishes anonymized speech,
unbeknownst to the attacker who still uses non-anonymized speech for enrollment.
Finally, in the {\it Semi-Ignorant} scenario, both the user and the attacker
use anonymized speech for publication and enrollment respectively. However the
parameters of anonymization used by the attacker might differ from the user's
parameters.
 
The final scenario is the one in which the user is most vulnerable, hence it is
considered as the lower bound for privacy in the context of this study.
Note that there
can be even stronger attacks \cite{srivastava2019evaluating} when the attacker
has the exact knowledge of the anonymization parameters and uses it to
generate large amounts of training data. This scenario is referred to in 
\cite{srivastava2019evaluating} as the {\it Informed} scenario. However it is
not very realistic, so we do not consider it here.

\subsubsection{Metrics}
\label{sec:metrics}
In all scenarios, the attacker implements the attack using a pretrained x-vector-PLDA based Automatic Speaker Verification ($ASV_\text{eval}$) system. Privacy protection is assessed in terms of the rate of failure of the attacker, as measured by the EER. The EER is computed from the distribution of PLDA scores generated by $ASV_\text{eval}$.
In addition, a pretrained Automatic Speech Recognition ($ASR_\text{eval}$) system is
used to decode anonymized speech and compute the WER for
utility evaluation. Both evaluation systems are trained on disjoint data from that used to train the anonymization system. For more details, see \cite{tomashenko:hal-02562199}.

Although we use Kaldi \cite{povey2011kaldi} to implement $ASV_\text{eval}$, we do not use it to compute the EER. Instead we
use the PLDA scores output by $ASV_\text{eval}$ as inputs to the cllr toolkit\footnote{\url{https://gitlab.eurecom.fr/nautsch/cllr}} to compute the ROCCH-EER \cite{tuprints9199}. The ROCCH-EER has interesting properties from the privacy perspective \cite{brummer2010measuring}. 
Its value does not exceed 50\% which is considered as the upper-bound for anonymization since it implies complete overlap between genuine and impostor PLDA score distributions \cite{gomez2017general}. The higher the ROCCH-EER and the lower the WER, the better.

\section{Experimental results}
\label{sec:exp-results}

\begin{figure}[t]
\centering
\includegraphics[width=\linewidth,trim=10 10 10 10,clip]{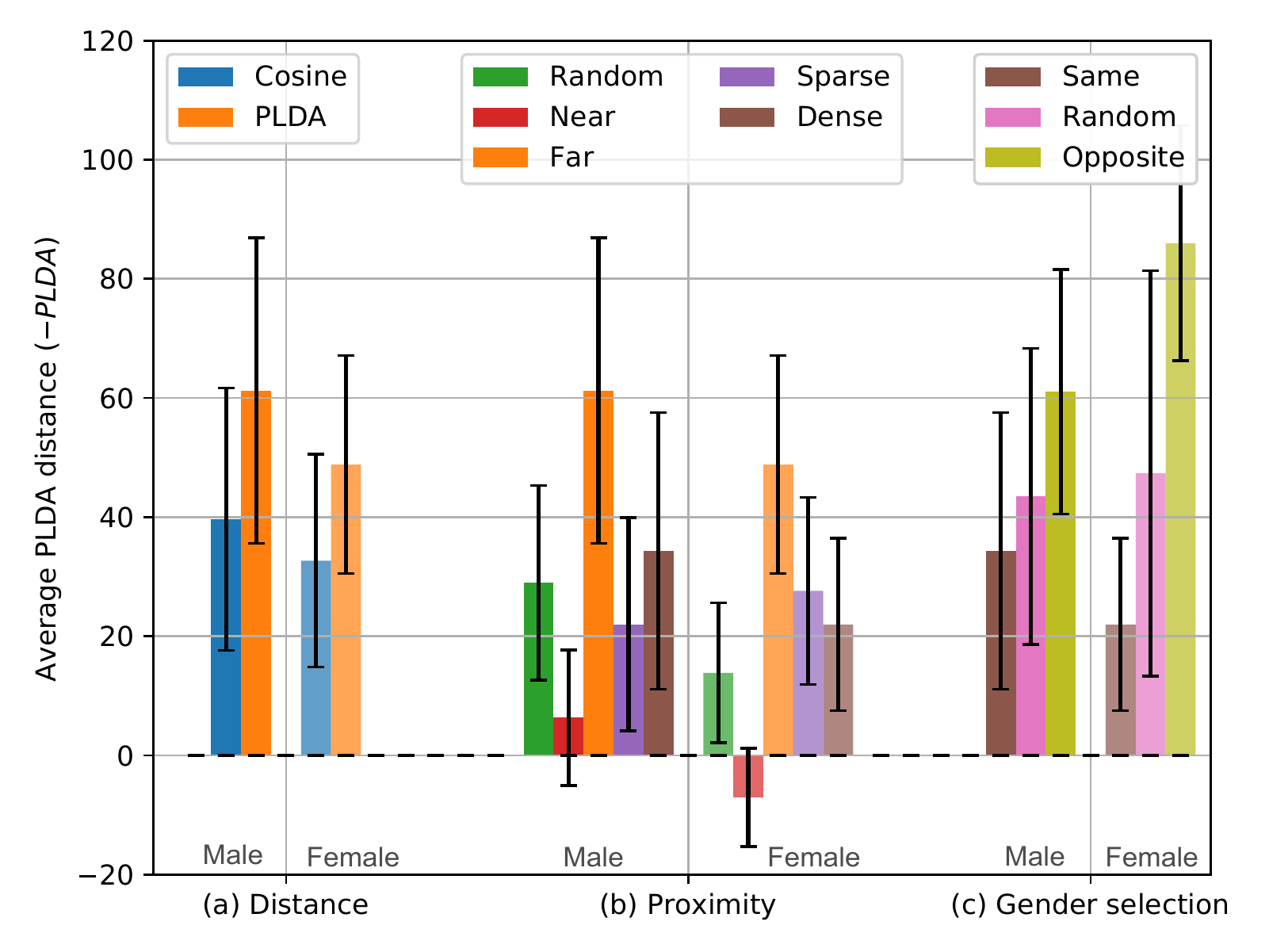}
\caption{Average PLDA distance between original and anonymized x-vectors for different design choices. Comparison of: a) Distance with proximity as \emph{far} and gender as \emph{same}; b) Proximity with distance as \emph{PLDA} and gender as \emph{same}; c) Gender with distance as \emph{PLDA} and proximity as \emph{dense}. (Darker left bars: male speakers, Lighter right bars: female speakers)}
\label{fig:distance_bars}
\end{figure}

All the experiments are performed using the publicly available recipe of the
VoicePrivacy Challenge.\footnote{
\url{https://github.com/Voice-Privacy-Challenge/Voice-Privacy-Challenge-2020}}
Figure~
\ref{fig:eer_bars} shows the EER values achieved by the considered anonymization scheme for different design choices. The corresponding WERs are reported in Table~\ref{tab:results_asr}. To qualitatively analyze the effect of
anonymization over the source speakers' x-vectors, we also compute the average PLDA distance
between original and anonymized x-vectors over all trial utterances in the
{\it test} set. 
Figure~\ref{fig:distance_bars} shows the average PLDA distance obtained 
for different design choices.

\begin{table}[t]
\caption{WER (\%) obtained by \emph{$ASR_\text{eval}$} on the dev and test sets.}
\label{tab:results_asr}
\centering
\footnotesize
\begin{tabular}{|c|c|c|c|c|}
\hline
\textbf{Distance}     & \textbf{Proximity}     & \textbf{\begin{tabular}[c]{@{}c@{}}Gender-\\ selection\end{tabular}} & \textbf{\begin{tabular}[c]{@{}c@{}}Dev \\ WER (\%)\end{tabular}} & \textbf{\begin{tabular}[c]{@{}c@{}}Test \\ WER (\%)\end{tabular}} \\ \hline \hline
\multicolumn{3}{|c|}{Baseline (no anonymization)}    & 3.83                  & 4.15                   \\ \hline \hline
\multicolumn{2}{|c|}{Random}     & \multirow{5}{*}{}                                                & 6.28                  & 6.58                   \\ \cline{1-2} \cline{4-5}
Cosine                & \multirow{2}{*}{Far}   & \multirow{5}{*}{Same}                                                & 6.50                  & 6.81                   \\ \cline{1-1} \cline{4-5} 
\multirow{6}{*}{PLDA} &                        &                                                                      & 6.38                  & 6.71                   \\ \cline{2-2} \cline{4-5} 
                      & Near                   &                                                                      & 6.42                  & 6.79                   \\ \cline{2-2} \cline{4-5} 
                      & Sparse                 &                                                                      & 10.04                 & 10.94                  \\ \cline{2-2} \cline{4-5} 
                      & \multirow{3}{*}{Dense} &                                                                      & 6.45                  & 6.83                   \\ \cline{3-5} 
                      &                        & Random                                                               & 6.86                  & 6.88                   \\ \cline{3-5} 
                      &                        & Opposite                                                             & 7.22                  & 7.19                   \\ \hline
\end{tabular}
\end{table}


\subsection{Distance}
Our first experiment aims to identify the distance metric which is most suitable for the selection of candidate target x-vectors. To do so, we fix the {\it proximity} as {\it far} and the gender selection strategy as {\it same}, and we consider cosine distance vs.\ PLDA. We observe in Fig.~\ref{fig:eer_distance} that cosine distance and PLDA result in a comparabley high ROCCH-EER in the {\it Ignorant} case but PLDA consistently outperforms cosine distance (i.e., it results in a higher ROCCH-EER) in the {\it Semi-Ignorant} case. We also notice in Fig.~\ref{fig:distance_bars} that the average PLDA distance between original and anonymized x-vectors is lower with cosine distance as compared to PLDA. For these reasons, we use PLDA to measure distances in x-vector space in the following experiments.


\subsection{Proximity}
Our second experiment assesses the five choices of target {\it proximity} described in Sections~\ref{sec:dc_proximity_random}, \ref{sec:dc_proximity_nf} and~\ref{sec:dc_proximity_sd}. The distance metric is fixed to PLDA and the gender selection strategy to {\it same}. We observe in Fig.~\ref{fig:eer_proximity} that although x-vector selection from a {\it far} region achieves the greatest level of anonymization in the {\it Ignorant} case, it is outperformed by selection from {\it sparse} or {\it dense} regions in the {\it Semi-Ignorant} case. We notice in Fig.~\ref{fig:distance_bars} that the target x-vectors are not too far from the source in the case of {\it sparse} or {\it dense} when compared to {\it far}. This may be due to the fact that \emph{same} gender selection allows only same-gender clusters which lie nearby the source x-vectors. {\it Random} target selection provides similar privacy protection and average PLDA distance as {\it sparse} or {\it dense}. 

Although {\it random} target selection produces comparable privacy protection and utility to {\it dense}, it limits the flexibility to select different regions in x-vector space. Compared to the {\it sparse} selection strategy, the {\it dense} strategy provides slightly better privacy protection in the {\it Semi-Ignorant} case, as well as higher utility (see Table~\ref{tab:results_asr}). This might be due to fewer members in sparse clusters, hence a smaller value of $N^*$ as pointed out in Section~\ref{sec:dc_proximity_nf}. Consequently we select the {\it dense} strategy in our third experiment.

\subsection{Gender selection}
Our third experiment concerns the gender selection strategy in Section~\ref{sec:dc_gs}. The distance is fixed to PLDA and proximity to {\it dense}. When we look at male trials in Fig.~\ref{fig:eer_gender}, it is not clear which {\it gender selection} strategy is the best among {\it same} and {\it opposite}, but female trials show that {\it random} strategy outperforms the rest. We also observe in Fig.~\ref{fig:distance_bars} that the mean distance is much higher in the case of {\it random} and {\it opposite} gender selection, which is intuitive since it allows selection of {\it dense} clusters from other genders as well. However, we notice that utility suffers in the case of {\it opposite} gender selection (see Table~\ref{tab:results_asr}) due to limitations of cross-gender voice conversion. Hence we can conclude that {\it random} gender selection is the best choice.


\section{Conclusions}
\label{sec:conc}

We presented a flexible speaker anonymization scheme as the primary baseline for the first VoicePrivacy Challenge. In particular we proposed three design choices for target selection in x-vector space, namely {\it distance metric}, {\it proximity}, and {\it gender selection} which can be combined to obtain various anonymization systems. We objectively evaluated these choices in terms of ROCCH-EER to measure privacy protection and decoding WER to measure utility. We also reported the average PLDA distance between the source and the target. We showed that the previously used cosine distance is not the best choice of distance in x-vector space and it should be replaced by PLDA. Then we explored interesting regions in the x-vector space for picking the target {\it pseudo-speaker} during anonymization. We observed that when the target is picked in a dense region and the target gender is selected at random, robust privacy protection can be achieved against both {\it Ignorant} and {\it Semi-Ignorant} attackers with a reasonable loss of utility. In the future, we will evaluate the best design choices with additional utility metrics, e.g., the WER obtained after retraining $ASR_\text{eval}$ on anonymized data.

\section{Acknowledgments}
This work was supported in part by ANR and JST under projects DEEP-PRIVACY, HARPOCRATES, and VoicePersonae, and by the European Union’s Horizon 2020 Research and Innovation Program under Grant Agreement No.\ 825081 COMPRISE (\url{https://www.compriseh2020.eu/}). Experiments presented in this paper were partially carried out using the Grid’5000 testbed, supported by a scientific interest group hosted by Inria and including CNRS, RENATER and several Universities as well as other organizations (see \url{https://www.grid5000.fr}).

\bibliographystyle{IEEEtran}

\bibliography{mybib}

\end{document}